\documentclass[12pt,psfig,a4]{amsart}

\usepackage{geometry}
\usepackage{amssymb}
\newtheorem{theorem}{Theorem}[section]
\newtheorem{lemma}[theorem]{Lemma}
\newtheorem{pro}[theorem]{Proposition}
\newtheorem{cor}[theorem]{Corollary}

\theoremstyle{definition}

\theoremstyle{remark}

\numberwithin{equation}{section}

\begin{document}
\bibliographystyle{plain}
\pagestyle{plain}
\pagenumbering{arabic}

\title{On Ruelle transfer operators and completely positive maps}

\author{Carlos F. Lardizabal}
\address{}
\curraddr{Instituto de Matem\'atica - Universidade Federal do Rio Grande do Sul - UFRGS. Av. Bento Gon\c calves 9500 - CEP 91509-900 - Porto Alegre - RS - Brazil}
\email{cfelipe@mat.ufrgs.br}
\thanks{}

\thanks{}



\date{}

\dedicatory{}

\begin{abstract}
We consider applications of transfer operators (also known as Ruelle operators) to completely positive maps (CPT) in quantum information theory. It is described a correspondence between fixed points of CPT maps and certain Markov-invariant measures. We also obtain a definition of entropy induced by transfer operators and an exponential decay property for mixed-unitary channels.
\end{abstract}

\maketitle

\def\laa{\langle}
\def\raa{\rangle}
\def\bv{\big\vert}
\def\SP{\hspace{1.5cm}}
\def\sp{\hspace{0.2cm}}
\def\qed{\begin{flushright} $\square$ \end{flushright}}
\def\qee{\begin{flushright} $\Diamond$ \end{flushright}}
\def\ov{\overline}

\section{Introduction}

Completely positive maps (CP) are of fundamental importance in describing open quantum dynamics such as quantum computations subject to noise \cite{bcs,nielsen,terhal}. Also known as quantum channels, CP maps are usually assumed to be trace-preserving (CPT) so they act on the space of density operators associated to a Hilbert space. Due to the fact that its eigenvalues are of norm less or equal to one, such maps are seen to exhibit a non-expansive behavior with respect to the trace norm \cite{ruskai2}. This also occurs with respect to $L^p$-norms if, moreover, the CPT map is identity-preserving \cite{ruskaipetz}.

\bigskip

An important result on quantum channels over a finite-dimensional Hilbert space is that a density operator is a fixed point for a CPT map $\Phi$ if and only if it is the barycenter of a measure which is invariant for the Markov operator associated to $\Phi$ \cite{lozinski,slomcz}. Such Markov maps, acting on measures, are dual to objects known as transfer operators \cite{jiang}. Transfer operators have been extensively studied in the theory of differentiable dynamical systems and statistical mechanics \cite{baladi,par}. Ruelle was among the first to use them to prove rigorous mathematical facts on thermodynamics, several of which has applications on expansive and hyperbolic systems. A basic transfer operator is obtained as follows: for some set $X$ consider $f:X\to X$ a map such that $f^{-1}(x)$ is countable or finite for each $x\in X$, and $g:X\to\mathbb{C}$ a function such that for each $x\in X$ the sum $\sum_{y\in f^{-1}(x)}g(y)$ is convergent. Then we define on functions $\psi:X\to\mathbb{C}$,
\begin{equation}
L_g(\psi)(x):=\sum_{f(y)=x} g(y)\psi(y).
\end{equation}

In connection to quantum mechanics one may proceed as follows. Let $D(\mathbb{C}^N)$ be the space of density operators over $\mathbb{C}^N$. One correspondence between quantum channels $\Phi$ and operators $T_\Phi:C(D(\mathbb{C}^N))\to C(D(\mathbb{C}^N))$ acting on continuous functions $\psi$ is
\begin{equation}
\Phi(\rho):=\sum_{i=1}^k p_iU_i\rho U_i^* \Rightarrow T_\Phi\psi(\rho):=\sum_{i=1}^k p_i\psi(U_i\rho U_i^*)
\end{equation}
Above, suppose that $\sum_i p_i=1$, $p_i\geq 0$ and that the $U_i$ are unitary. We say that $\Phi$ is a mixed-unitary channel. Another possible correspondence is
\begin{equation}
\Phi(\rho):=\sum_{i=1}^k p_iU_i\rho U_i^* \Rightarrow T_\Phi'\psi(\rho):=\sum_{i=1}^k \psi(p_iU_i\rho U_i^*)
\end{equation}
Note that $T_\Phi$ and $T_\Phi'$ are the same only when $\psi$ is linear. We say that $T_\Phi$ and $T_\Phi'$ are transfer operators associated to the quantum channel $\Phi$. In a non-commutative setting, one can also study transfer operators associated to quantum spin chains \cite{matsui,ogata}.

\bigskip

A general question one may ask is whether transfer operators have interesting connections with completely positive  dynamics. In particular one might attempt to modify classical methods to produce new proofs on CPT maps. However, such channels do not present expansive or hyperbolic behavior, so first of all one has to determine how the methods employed in the known classical settings can be adapted to quantum dynamics. We show that some of these adaptations are indeed possible, serving as an example of how a classical tool can be used to aid the description of quantum problems.

\bigskip

The goals of the present work, all related to some kind of transfer operator, are the following:

\bigskip

a) {\bf The barycenter theorem}. In \cite{lozinski} it was stated that in a finite-dimensional Hilbert space a density operator $\rho_0$ is a fixed point for a CPT map $\Phi$ if and only if it is the barycenter of a measure $\mu_0$ which is invariant for the Markov operator $P_\Phi$ associated to $\Phi$:
\begin{equation}
\Phi(\rho_0)=\rho_0 \Longleftrightarrow \;\;\psi(\rho_0)=\int \psi(\rho)\;d\mu_0(\rho),\;\;\;\forall\psi\in D(\mathbb{C}^N)^*, \;\; P_\Phi(\mu_0)=\mu_0.
\end{equation}
We call this result the barycenter theorem. The work \cite{slomcz} contains a proof of this fact in a much more general context than finite-dimensional quantum theory. For completeness, we describe such demonstration, which we believe is not well-known in the physics community, emphasizing the details which are relevant to our purposes. Later we describe applications related to the barycenter theorem.

\bigskip

Proofs involving barycenters of measures are present in the theory of decomposition of states, where one can find correspondences between spaces of functions and certain operator subalgebras. An example is given by Tomita's theorem which states that an orthogonal $\mu$ measure over the states of an algebra $\mathcal{A}$ implies an isomorphism between maps in $L^{\infty}(\mu)$ and abelian subalgebras of certain representations of $\mathcal{A}$; conversely, such isomorphism implies the orthogonality of $\mu$ \cite{bratteli}. In particular this theorem shows that orthogonal measures are extremal in the set of measures which have a common barycenter. Basic references on probability measures over the space of density matrices are \cite{bengts,sanpera}.

\bigskip

b) {\bf Entropies induced by transfer operators.} As an application we give a definition of dynamical entropy induced by a quantum channel, and prove some of its properties. We define
\begin{equation}
h_Q(\rho):=-\sum_i tr(Q_i\rho)\sum_j tr(Q_jU_i\rho U_i^*)\log tr(Q_jU_i\rho U_i^*)
\end{equation}
where $\sum_i Q_i=I$ are positive matrices. Note that if the $Q_i$ are multiples of the identity matrix, the above  definition reduces to Shannon entropy. A similar object has been described in \cite{BLLT}, in a slightly different setting, where a version of the variational principle of pressure for density matrices is proved, along with a restatement of the Holevo bound, both in terms of a quantum channel-induced entropy formula. In that matter, it is natural to consider mappings which are more general than CPT maps, so there follows a brief discussion on channels with place-dependent probabilities, so called nonlinear channels. Again, in this setting, our definition of entropy is obtained from a transfer operator. Such entropy allows one to obtain a relation involving the entanglement of formation,
\begin{equation}
E(\omega)=\inf_{\mu\in M_\omega(E_{\mathbb{C}^N})}\int S\circ r(\eta)\;d\mu(\eta),
\end{equation}
as described in \cite{majewski} and which we briefly discuss in an example.

\bigskip

c) {\bf Exponential mixing property for transfer operators.} It is well-known that CPT maps with a unique fixed point present exponential convergence,
\begin{equation}
\Vert \Phi(\rho)^n-\rho_0\Vert_{tr}\leq C_N\vert \kappa\vert^n
\end{equation}
where $\vert \kappa\vert$ is second largest absolute value among the eigenvalues of $\Phi$ \cite{terhal}. Motivated by this fact, one might expect that the successive iterates of some associated transfer operator behave in a similar way. In section \ref{mixingsection} we describe the problem of finding fixed points for a transfer operator associated to mixed-unitary channels. This result is used to prove an exponential mixing property. More precisely, we prove:

\bigskip

{\bf Theorem (Exponential decay).} Let $\Phi=\sum_i p_iF_i$ be a mixed-unitary quantum channel, and let $T_\Phi\varphi(\rho)=\sum_i \varphi(p_iF_i(\rho))$ the associated transfer operator. Given $\nu$-H\"older continuous functions  $\varphi$, $\psi:D(\mathbb{C}^N)\to\mathbb{R}$, there is $K=K(\varphi_1,\varphi_2)$ such that
\begin{equation}
\Bigg\arrowvert\int \psi(T_\Phi^n\varphi)\; dm - \int\psi\varphi_0\; dm\int\varphi \;dm\Bigg\arrowvert\leq K\Lambda_1^n, \;\; n\geq 0,
\end{equation}
where $\Lambda_1=1-e^{-D_1}$, $D_1=\sup\{\theta(\varphi_1,\varphi_2):\varphi_1,\varphi_2\in C(\lambda_1a,\nu)\}$, $a>0$, $\nu >0$ and $\varphi_0$ is the fixed point for $T_\Phi$.

\bigskip

Clearly, a proof showing the existence of $\varphi_0$ is required. Also the constant $\Lambda_1$ is obtained from the finiteness of the transfer operator with respect to Hilbert's projective metric. This will be made precise and proved later. Recently \cite{wolf} the projective metric has been used to prove that certain CPT maps are strictly contractive by improving the well-known result with respect to the trace norm, $\Vert\Phi(\rho)-\Phi(\eta)\Vert_1\leq \Vert\rho-\eta\Vert_1$, by stating
\begin{equation}\label{tanheq}
\Vert\Phi(\rho)-\Phi(\eta)\Vert_{tr}\leq \tanh\Big(\frac{\Delta(\Phi)}{4}\Big)\Vert\rho-\eta\Vert_{tr},
\end{equation}
where $\Delta(\Phi):=\sup\{\theta(\Phi(A_1),\Phi(A_2)): A_1, A_2\in P(\mathbb{C}^N)\}$, and $P(\mathbb{C}^N)$ denotes the positive matrices.  The number $\Delta(\Phi)$ is the projective diameter of (the image of) $\Phi$ and a main point is to determine when it is finite. The proof of the exponential decay is motivated by a construction made for expansive/hyperbolic dynamics, where Ruelle's transfer operator has a central role in the entire argument \cite{baladi},\cite{viana}. In order to follow this program for transfer operators associated to CPT maps, we have to make some adaptations. An important part of our reasoning is to find a fixed point for the associated transfer operator, obtained by verifying the hypothesis found in the work \cite{jiang} which, in our case, reduces to
\begin{equation}
\sup_{\rho\in D(\mathbb{C}^N)}\sum_{\vert J\vert=k}\gamma_J(\rho)<r(T_\Phi)^k,
\end{equation}
where $r(T_\Phi)$ is the spectral radius of $T_\Phi$ and
\begin{equation}
\gamma_J(\rho)=\sup_{\rho\neq\eta}\frac{\Vert F_J(\rho)-F_J(\eta)\Vert}{\Vert\rho-\eta\Vert}.
\end{equation}
If the above is true for a given non-expansive system, for some $k$, then Ruelle's theorem is valid and, in particular, a fixed point for the transfer operator exists. We believe that it is possible to prove the existence of the fixed point by other means (via projective metric-related arguments, for instance), but we decided to use the above result because of its natural connection to the kind of dynamics treated in this paper.

\bigskip

An outline of this work is as follows. In section \ref{snotpre} we review some basic facts and notations on CPT maps and transfer operators, including Ruelle's theorem. In section \ref{sec_baricent} we give a description of the barycenter theorem, following \cite{slomcz}. In section \ref{sec_entrop_ex} we give an application of transfer operators by defining a dynamical entropy induced by a quantum channel. We establish a simple relation of such entropy with the entanglement of formation, described in \cite{majewski}. In section \ref{sec_pmetric} we review basic facts on the projective metric and in section \ref{mixingsection} we prove an exponential decay theorem for mixing-unitary channels. We remark that some parts of what is presented is not specific of completely positive maps. It is clear from the context whether such maps could be replaced for positive ones, for instance.

\section{Notations and preliminaries}\label{snotpre}

\bigskip

Let $M_N(\mathbb{C})$ be the set of order $N$ matrices with the Hilbert-Schmidt inner product $\langle A,B\rangle=tr(A^*B)$. Recall that an operator $\rho$ on $M_N(\mathbb{C})$ is positive, denoted $\rho\geq 0$ if it is hermitian and has nonnegative eigenvalues (i.e. its matrix is positive semidefinite). Let $D(\mathbb{C}^N)=\{\rho\in M_N(\mathbb{C}): \rho\geq 0, \;tr(\rho)=1\}$ be the space of {\bf density matrices} over $\mathbb{C}^N$. It is a compact, convex subset of $M_N(\mathbb{C})$.  We can identify the space of state functionals $E_{\mathbb{C}^N}=\{\omega:M_N(\mathbb{C})\to\mathbb{C}: \omega\geq0,\omega(I)=1\}$ with $D(\mathbb{C}^N)$, since every state $\omega$ can be written as $\omega(X)=tr(X\rho_\omega)$, where $\rho_\omega\in D(\mathbb{C}^N)$ is uniquely determined.
Consider the operator $\Lambda:D(\mathbb{C}^N)\to D(\mathbb{C}^N)$,
\begin{equation}
\Lambda(\rho):=\sum_{i=1}^k F_i(\rho)
\end{equation}
A particular example of interest is when
\begin{equation}\label{muop}
F_i(\rho)=p_iU_i\rho U_i^*,
\end{equation}
where $\sum_{i=1}^k p_i=1$, $p_i\geq 0$, and the $U_i$ are unitary matrices. In this case we call $\Lambda$ a random unitary operator or a {\bf mixed-unitary operator}. More generally, any operator $\Phi$ which can be written in the form
\begin{equation}
\rho\stackrel{\Phi}{\mapsto}\sum_{i=1}^k V_i\rho V_i^*
\end{equation}
for $V_i$ linear maps, is said to be {\bf completely positive}. The above form is the Kraus representation and any two such representations are unitarily equivalent (i.e., equal up to a unitary matrix). If $\sum_i V_i^*V_i=I$, then the operator is {\bf trace-preserving}, $tr(\Phi(\rho))=tr(\rho)$ and if $\sum_i V_iV_i^*=I$ then the operator is {\bf unital}, that is, $\Phi(I)=I$. Note that mixed-unitary operators are trace-preserving and unital. We refer the reader to the literature for more details on such maps \cite{nielsen,petzbook}. In this work we shall mainly consider CPT maps which are mixed-unitary, but more general channels will be discussed as well.

\bigskip

A {\bf transfer operator} is an operator of the form $T:C(D(\mathbb{C}^N))\to C(D(\mathbb{C}^N))$,
$T\varphi(\rho):=\sum_i G_i(\varphi,\rho)$ for some functions $G_i$, which are not necessarily linear. We are interested in certain operators which have appeared in the theory of dynamical systems, iterated function systems, and we wish to study others which may provide applications to quantum information theory. Below we list three basic examples.

\bigskip

a) {\bf Ruelle transfer operator}. For the shift operator $\sigma$ on the space $\Omega=\{1,\dots,k\}^{\mathbb{N}}$, one can define on the space of continuous (H\"older) functions the Ruelle transfer operator,
\begin{equation}
L\phi(x)=\sum_{\sigma(y)=x} e^{f(y)}\phi(y).
\end{equation}

\bigskip

b) {\bf Contractive transfer operator}. Let $\varphi\in C(D(\mathbb{C}^N))$ where $C(D(\mathbb{C}^N))$ denotes the set of real continuous functions defined on the density matrices. We could be more general and consider the set of measurable functions instead, but for simplicity we consider continuous functions. Define $T_c:C(D(\mathbb{C}^N))\to C(D(\mathbb{C}^N))$ as
\begin{equation}
T_c\varphi(\rho):=\sum_{i=1}^k \varphi(p_iF_i(\rho)).
\end{equation}
In general, we assume that $p_i\geq 0$, $\sum_i p_i=1$, and that the $U_i$ are unitary matrices. In this way, we have a transfer operator induced by a mixed-unitary channel.
\bigskip

c) {\bf Barycentric transfer operator}. Define $T_b: C(D(\mathbb{C}^N))\to C(D(\mathbb{C}^N))$ as
\begin{equation}
T_b\varphi(\rho):=\sum_{i=1}^k p_i\varphi(F_i(\rho)), \;\;p_i\geq 0,\;\;\sum_i p_i=1.
\end{equation}

Denote $\langle \varphi,\mu\rangle=\int\varphi\;d\mu$ and let $M^1(D(\mathbb{C}^N))$ denote the set of probability measures over $D(\mathbb{C}^N)$. Define the {\bf Markov operator} $P_b:M^1(D(\mathbb{C}^N))\to M^1(D(\mathbb{C}^N))$ on measures associated to $T_b$,
\begin{equation}
P_b\mu(E):=\sum_{i=1}^k p_i\int_{F_i^{-1}(E)} d\mu=\sum_i p_i\mu(F_i^{-1}(E)),
\end{equation}
for any Borel set $E$.
It is a simple matter to prove that $P_b=T_b^*$, that is, $\langle T_b\varphi,\mu\rangle=\langle\varphi,P_b\mu\rangle$, for any choice of $\mu\in M^1(D(\mathbb{C}^N))$, $\varphi\in C(D(\mathbb{C}^N))$. In the literature, it is also said that $P_b$ is a Feller operator with conjugate $T_b$.
We will need in the next section the following fact concerning $T_b$. The proof is straightforward.

\begin{lemma}\label{altpropp}
For any CPT map $\Lambda(\rho)=\sum_i p_iF_i(\rho)$, $\Psi\circ \Lambda=T_b\circ\Psi$, for all $\Psi\in D(\mathbb{C}^N)^*$.
\end{lemma}

Recall that $D(\mathbb{C}^N)^*$ denotes the linear functionals over $D(\mathbb{C}^N)$. For reference, we state a version of the well-known Ruelle theorem, used in \cite{jiang}. This will be necessary in section \ref{mixingsection}.

\bigskip

{\bf Definition.} Let $(X,\{w_j\},\{q_j\})_{j=1}^k$ be a {\bf non-expansive system}, that is, $|w_j(x)-w_j(y)|\leq |x-y|$ for all $x,y\in X$ and all $j$. We assume the maps $w_j$ act on the compact set $X\subset\mathbb{R}^d$, for some $d$, and the $q_i$ are nonnegative functions on $X$. Note that in the definition of non-expansive system it is not necessarily assumed that $\sum_i q_i(x)=1$. Let $K\subset X$ be the unique nonempty compact which is invariant under $\{w_i\}_{j=1}^k$. We say that the {\bf Ruelle operator theorem} holds for this system if there exists a unique positive function $h\in C(K)$ and a unique probability measure $\mu\in M(K)$ such that $Th=\lambda h$, $T^*\mu=\lambda\mu$, $\langle h,\mu\rangle=1$, and for every $f\in C(K)$, $\lambda^{-n}T^n f$ converges to $\langle f,\mu\rangle h$ in the supremum norm.

\bigskip

The number $\lambda=\lim_n\Vert T^n \Vert$ is the spectral radius of the transfer operator $T$, given by
\begin{equation}
T\varphi(x)=\sum_{i=1}^k p_j(x)\varphi(w_j(x))
\end{equation}
The description of a mixed-unitary CPT map is a particular example of the above setting. In that case the set $X$ equals a bounded cone of positive matrices and we take $w_i(\rho)=F_i(\rho)=p_iU_i\rho U_i^*$, with all $q_i$ equal to 1. Note that for the case of one qubit, a mixed-unitary $\Phi$ can be seen to act on the unit sphere of $\mathbb{R}^3$ \cite{kingruskai}, but the contractive parts of the associated transfer operator (the $w_i$ maps) act on a (bounded) cone of matrices, hence a strictly larger set.

\section{Barycenter theorem}\label{sec_baricent}

Now we consider sequences of measures in $D(\mathbb{C}^N)$. Mathematical foundations of the theory can be found in \cite{bills}. Let $M^1=M^1(D(\mathbb{C}^N))$ denote the set of probability measures. Since $D(\mathbb{C}^N)$ is compact we have that every sequence of probability measures is tight. By Prokhorov's theorem, the sequence is relatively compact in the topology of weak convergence. Let $(\mu_n)_{n\in\mathbb{N}}$ be a sequence of probability measures. The generalized limit of the sequence $(\mu_n)_{n\in\mathbb{N}}$ is defined as
\begin{equation}
Lim(\mu_n)_{n\in\mathbb{N}}:=\overline{co}\{\nu\in M^1: \nu \textrm{ is as accumulation point of } (\mu_n)_{n\in\mathbb{N}}\}
\end{equation}
It is a compact and convex subset of $M^1$. Also, we have that
$$Lim(\mu_n)_{n\in\mathbb{N}}=\{\nu\in M^1:\forall g\in C(D(\mathbb{C}^N)), \;\;\liminf_{n\to\infty}\langle g,\mu_n\rangle\leq\langle g,\nu\rangle\}$$
\begin{equation}
=\{\nu\in M^1:\forall g\in C(D(\mathbb{C}^N)), \;\;\langle g,\nu\rangle\leq \limsup_{n\to\infty}\langle g,\mu_n\rangle\}
\end{equation}
If $\mu_n\stackrel{w^*}{\rightarrow}\nu$ as $n\to\infty$ then $Lim(\mu_n)_{n\in\mathbb{N}}=\{\nu\}$.

\bigskip

{\bf Definition.} We say that $P:M^1\to M^1$ is a {\bf Markov operator} on probability measures if $P(\lambda\mu_1+(1-\lambda)\mu_2)=\lambda P\mu_1+(1-\lambda)\mu_2$, for all $\mu_1, \mu_2\in M^1$, and $\lambda\in(0,1)$.

\bigskip

If $P:M^1\to M^1$ is a Markov operator, we say that a measure is {\bf $P$-persistent} if the sequence $(P^n\mu)_{n\in\mathbb{N}}$ is relatively compact. Since we are considering density matrices in a finite-dimensional vector space, we have that every $\mu\in M^1$ is $P$-persistent. Also if $P$ is Markov, we say that $\mu\in M^1$ is {\bf $P$-invariant} if $P\mu=\mu$. The set of all such measures is denoted by $M^P=M^P(D(\mathbb{C}^N))$. It is a convex subset of $M^1$. Finally, define
\begin{equation}
S(\mu):=M^P(D(\mathbb{C}^N))\cap Lim(P^n\mu)_{n\in\mathbb{N}}
\end{equation}
Recall that a Markov operator is Feller if there exists $T:C(D(\mathbb{C}^N))\to C(D(\mathbb{C}^N))$ such that $\langle Tf,\mu\rangle=\langle f,P\mu\rangle$, for all $f\in C(D(\mathbb{C}^N))$ and $\mu\in M^1$. For CPT maps, if $P$ is Feller then $S(\mu)$ is a non-empty compact and convex subset of $M^1$.

\begin{theorem}
Let $P:M^1\to M^1$ be a Feller operator, let $\mu\in M^1$ and $g\in C(D(\mathbb{C}^N))$. Then
\begin{enumerate}
\item $$\lim_{n\to\infty}\Big[\limsup_{j\to\infty}\frac{1}{n}\sum_{i=0}^{n-1}\langle g,P^{i+j}\mu\rangle\Big]=\inf_{n\in\mathbb{N}} \Big[\limsup_{j\to\infty}\frac{1}{n}\sum_{i=0}^{n-1}\langle g,P^{i+j}\mu\rangle\Big]$$
    $$=\max\{\langle g,\nu\rangle: \nu\in S(\mu)\}$$
Also, $\limsup_j$ can be replaced by $\sup_{j\in\mathbb{N}}$.
\item $$\lim_{n\to\infty}\Big[\liminf_{j\to\infty}\frac{1}{n}\sum_{i=0}^{n-1}\langle g,P^{i+j}\mu\rangle\Big]=\sup_{n\in\mathbb{N}} \Big[\liminf_{j\to\infty}\frac{1}{n}\sum_{i=0}^{n-1}\langle g,P^{i+j}\mu\rangle\Big]$$
    $$=\min\{\langle g,\nu\rangle: \nu\in S(\mu)\}$$
Also, $\liminf_j$ can be replaced by $\inf_{j\in\mathbb{N}}$.
\end{enumerate}
\end{theorem}
{\bf Proof.} The result follows by \cite{slomcz}, theorem 2.3, p. 46, since for density matrices in a finite-dimensional space every $\mu$ is $P$-persistent.

\qed

\begin{theorem}\cite{slomcz} Let $\Lambda=\sum_i p_iF_i$ be a CPT map. A density matrix $\rho_0$ is a fixed point for $\Lambda$ if and only if $\rho_0$ is the barycenter of some $\mu\in M^P$.
\end{theorem}
{\bf Proof.} Let $\rho_0$ be fixed for $\Lambda$, let $\Psi\in D(\mathbb{C}^N)^*$. We have $\Psi\circ\Lambda^n=T^n\circ \Psi$, by lemma \ref{altpropp}. Then
\begin{equation}
\Psi(\rho_0)=\Psi(\Lambda^n(\rho_0))=\int\Psi\circ\Lambda^n \;d\delta_{\rho_0}=\int T^n\Psi\;d\delta_{\rho_0}=\int\Psi\; dP^n\delta_{\rho_0}
\end{equation}
Write $n=i+j$ so we deduce that
\begin{equation}
\Psi(\rho_0)=\lim_{n\to\infty}\Big[\liminf_{j\to\infty}\frac{1}{n}\sum_{i=0}^{n-1}\langle \Psi,P^{i+j}\delta_{\rho_0}\rangle\Big]=\min\{\langle \Psi,\nu\rangle: \nu\in S(\delta_{\rho_0})\}
\end{equation}
\begin{equation}
\Psi(\rho_0)=\lim_{n\to\infty}\Big[\limsup_{j\to\infty}\frac{1}{n}\sum_{i=0}^{n-1}\langle \Psi,P^{i+j}\delta_{\rho_0}\rangle\Big]=\max\{\langle \Psi,\nu\rangle: \nu\in S(\delta_{\rho_0})\}
\end{equation}
Therefore $\Psi(\rho_0)=\int \Psi d\mu$ for any $\mu\in S(\delta_{\rho_0})$ and $\Psi\in D(\mathbb{C}^N)^*$. All $\mu$ in $S(\delta_{\rho_0})$ are $P$-invariant, by definition, so we are done. Conversely, assume that $\rho_0$ is the barycenter of some $\mu$ which is $P$-invariant. If $\Psi\in D(\mathbb{C}^N)^*$, so is $T\circ\Psi$. Then
\begin{equation}
\Psi(\Lambda(\rho_0))=T(\Psi(\rho_0))=\int T\circ \Psi\;d\mu=\int\Psi\;dP\mu=\int\Psi\;d\mu=\Psi(\rho_0)
\end{equation}
Since $\Psi$ is arbitrary, this implies that $\Lambda(\rho_0)=\rho_0$.
\qed

{\bf Remark.} The above proof also shows that if $\rho_0$ is a fixed point for $\Lambda$, then it is the barycenter of every $\mu\in S(\delta_{\rho_0})$.

\bigskip

\begin{cor}
If there is a unique $T$-invariant measure, where $T$ is the transfer operator conjugate to the Markov operator for measures, then the associated quantum channel admits a unique fixed point.
\end{cor}
{\bf Proof.} In fact, if $\rho_1$ is a fixed point, then by the theorem it is the barycenter of a $P$-invariant measure, say, $\mu_1$. But by assumption we must have $\mu_1=\mu_0$. Since the barycenter of a measure is unique \cite{bratteli}, we must have $\rho_1=\rho_0$.
\qed

\section{Example: entropy induced by transfer operators}\label{sec_entrop_ex}

Now we consider channels of the form $\Phi(\rho)=\sum_i p_i(\rho)G_i(\rho)$, that is, the $p_i$ are position-dependent. Such operators are not linear in $\rho$, so to make distinction from the ones we have been considering so far we call such objects {\bf nonlinear channels}. We assume that $\sum_i p_i(\rho)=1$, for all $\rho$, and that the $G_i$ are operators of the form $G_i(\rho)=U_i\rho U_i^*$, $U_i$ unitary. We have that $\Phi$ induces the transfer operator
\begin{equation}\label{top_entrqifs}
T'\varphi(\rho)=\sum_i p_i(\rho)\varphi(G_i),
\end{equation}
which satisfies $\langle T'\varphi,\mu\rangle=\langle \varphi, P'\mu\rangle$, where $P'\mu(E)=\sum_{i=1}^k\int_{F_i^{-1}(E)}p_i(\rho) d\mu$.

\bigskip

Let $\eta:\mathbb{R}^+\to\mathbb{R}$, $\eta(\rho):=-x\ln(x)$ if $x\neq 0$ and equal to zero otherwise. Define $h(\rho):=\sum_i \eta(p_i(\rho))$, where we set $p_i(\rho):=tr(W_i\rho W_i^*)$, for some $\{W_i\}_i$ satisfying $\sum_i W_i^*W_i=I$. For brevity we write $Q_i=W_i^*W_i$, so $p_i(\rho)=tr(Q_i\rho)$. Define $h_Q(\rho):=T'h(\rho)=\sum_i p_i(\rho)h(G_i(\rho))$. We write such expression more explicitly:
\begin{equation}
h_Q(\rho)=\sum_i tr(Q_i\rho)\sum_j \eta(p_j(G_i(\rho)))=-\sum_i tr(Q_i\rho)\sum_j tr(Q_jU_i\rho U_i^*)\log tr(Q_jU_i\rho U_i^*)
\end{equation}

We call $h_Q$ the {\bf transfer entropy} associated to $\Phi$. If the $Q_i$ are multiples of the identity matrix, the above definition reduces to Shannon entropy. We list some of its properties, together with a basic lemma on transfer operators. We say that a channel $\sum_i p_iF_i$ is {\bf homogeneous} if both $p_i$ and $p_iF_i$ are affine maps. Note that every linear quantum channel (i.e., all $p_i$ constant and $F_i$ linear) is homogeneous. Also, the barycenter theorem holds for homogeneous channels \cite{slomcz}.

\begin{pro}\label{homog_lem1}
Let $\Phi(\rho)=\sum_i p_i(\rho)G_i(\rho)$. The following are valid:
\begin{enumerate}
\item $0\leq h_Q(\rho)\leq \log k$.
\item For all $\alpha\in (0,1)$, and all $\rho_1,\rho_2\in D(\mathbb{C}^N)$,
$$h_Q(\alpha\rho_1+(1-\alpha)\rho_2)\geq \alpha^2 h_Q(\rho_1)+(1-\alpha)^2 h_Q(\rho_2).$$

If the channel is homogeneous then:

\item If $\varphi$ is concave then $T'\varphi$ is concave.
\item $h_Q^n(\rho)\leq h\circ \Phi(\rho)^n$, where $h$ is Shannon entropy.
\end{enumerate}
\end{pro}

{\bf Proof.} The proof of (3) and (4) are simple and can be seen in \cite{slomcz}. The proof of (1) is straightforward from the fact that $h_Q$ can be seen as a convex combination of Shannon entropies. For the proof of (2), note that we can write $h_Q(\rho)=\sum_i p_i(\rho) S_i(\rho)$
where $p_i$ represents the probabilities and $S_i(\rho)=-\sum_j \eta_j^i(\rho)\log \eta_j^i(\rho)$,
where
\begin{equation}
\eta_j^i(\rho)=tr(Q_jU_i\rho U_i^*)
\end{equation}
and these are positive with $\sum_j \eta_j^i=1$. Then
the lemma follows from the fact that each of the $S_i$ is
a concave function:
$$h_Q(\alpha\rho_1+(1-\alpha)\rho_2)=\sum_i p_i(\rho)S_i(\alpha\rho_1+(1-\alpha)\rho_2)\geq \sum_i p_i(\rho)(\alpha S_i(\rho_1)+(1-\alpha)S_i(\rho_2))$$
$$=\sum_i [\alpha p_i(\rho_1)+(1-\alpha)p_i(\rho_2)][\alpha S_i(\rho_1)+(1-\alpha)S_i(\rho_2)]$$
$$=\sum_i \alpha^2p_i(\rho_1)S_i(\rho_1) +\alpha(1-\alpha)\Big[p_i(\rho_1)S_i(\rho_2)+p_i(\rho_2)S_i(\rho_1)\Big]+(1-\alpha)^2p_i(\rho_2)S_i(\rho_2)\geq $$
\begin{equation}
\geq\sum_i \alpha^2p_i(\rho_1)S_i(\rho_1) +(1-\alpha)^2p_i(\rho_2)S_i(\rho_2)=\alpha^2 h_Q(\rho_1)+(1-\alpha)^2h_Q(\rho_2)
\end{equation}

\qed

Define the {\bf relative} transfer entropy of the (nonlinear) quantum channel $\Phi_A(\rho)=\sum_i tr(Q_i^A\rho)U_i^A\rho U_i^{A*}$ with respect to $\Phi_B(\rho)=\sum_i tr(Q_i^B\rho)U_i^B\rho U_i^{B*}$  by
\begin{equation}
h_Q(\Phi_A\vert\Phi_B)(\rho):=
\sum_{i=1}^k tr(Q_i^A\rho)\sum_{j=1}^k tr(Q_j^AU_i^A\rho U_i^{*A})\Big(\log tr(Q_j^AU_i^A\rho U_i^{*A})-\log tr(Q_j^BU_i^B\rho U_i^{*B})\Big)
\end{equation}
Note we can also write
\begin{equation}\label{klein_soma}
h_{Q}(\Phi_A\vert\Phi_B)(\rho)=\sum_{i=1}^k tr(Q_i^A\rho)h(\rho_i^A\vert\rho_i^B)
\end{equation}
where $h$ denotes Shannon entropy, $h(\cdot|\cdot)$ its relative version, $\rho_i^A$ is the diagonal matrix with entries $tr(Q_j^AU_i^A\rho U_i^{*A})$, $j=1,\dots, N$, and analogously for $\rho_i^B$. So we can take equation (\ref{klein_soma}) as our definition and from that, the following property holds, a consequence of Klein's inequality:

\bigskip

\begin{lemma}
a) If $tr(Q_i^A\rho_A)\neq 0$ for all $i$, then $h_{Q}(\Phi_A\vert\Phi_B)\geq 0$ and equality holds if and only if $\rho_i^A=\rho_i^B$ for all $i$.
b) The relative transfer entropy is jointly convex in its arguments, that is for all $0\leq \lambda\leq 1$, and operators $\rho_1, \rho_2, \eta_1, \eta_2$,
\begin{equation}
h_{Q}(\lambda \rho_1+(1-\lambda)\rho_2\;|\;\lambda \eta_1+(1-\lambda)\eta_2)\leq \lambda h_{Q}(\rho_1\;\vert \; \eta_1)+(1-\lambda)h_{Q}(\rho_2\;\vert\; \eta_2)
\end{equation}
\end{lemma}

\qed

{\bf Remark.} To see how the relative (von Neumann) entropy is related to the above, we proceed in the following way. For any fixed dynamics $U_i^A$ and $U_i^B$, we choose $Q_i^A$ and $Q_i^B$ so that the traces involving such mappings are constant. For instance if we write $Q_i^A=p_iI$ and $Q_i^B=q_iI$, where $I$ is the identity operator and $p_i$, $q_i$ are positive numbers with $\sum_i p_i=\sum_i q_i=1$, then (\ref{klein_soma}) reduces to
\begin{equation}
h_{Q}(\Phi_A\vert\Phi_B):=
\sum_{i=1}^k p_i\sum_{j=1}^k p_j\Big(\log p_j-\log q_j\Big)=\sum_j p_j\log p_j -\sum_j p_j\log q_j,
\end{equation}
so we recover the classical relative entropy as a particular case.

\bigskip

{\bf Example (entanglement of formation).} Let $\mu_0$ be the minimizing measure for
\begin{equation}
E(\omega):=\inf_{\mu\in M_\omega(E_{\mathbb{C}^N})}\int S\circ r(\eta)\; d\mu(\eta),
\end{equation}
where $r:D(\mathbb{C}^N)\otimes D(\mathcal{H})\to D(\mathbb{C}^N)$, for some Hilbert space $\mathcal{H}$, is the restriction
\begin{equation}
r\omega(A):=\omega(A\otimes I)
\end{equation}
The function $E$ is the entanglement of formation defined in \cite{majewski}. The existence of $\mu_0$ is evident, due to the continuity of $E$. The set of states with barycenter $\omega$ is denoted $M_\omega(E_{\mathbb{C}^N})$, where $E_{\mathbb{C}^N}$ is the set of states over $\mathbb{C}^N$.
Let $P$ be a Markov operator such that $P\mu_0=\mu_0$. In principle, it is not evident that a nontrivial $P$ should exist. If it does, we have an inequality involving the entanglement of formation of a state in terms of the transfer entropy associated to quantum channels. In fact, by the barycenter theorem, we have that $\rho_0$, the barycenter of $\mu_0$ satisfies $\Phi(\rho_0)=\rho_0$, where $\Phi$ is the quantum channel induced by $P$ (recall the correspondence between the density matrix $\rho_0$ and the state $\omega$). Then
\begin{equation}
E(\omega)=\langle S\circ r,\mu_0\rangle=\langle S\circ r,P\mu_0\rangle=\langle T'S\circ r,\mu_0\rangle,
\end{equation}
where $T'$ is given by (\ref{top_entrqifs}). Define $E_{h_Q}(\omega)=\inf_{\mu\in M_\omega(E)}\int h_Q^r(\eta)\; d\mu(\eta)$, where $h_Q^r(\rho)=h_Q(r\circ \rho)$. We conclude the following:

{\it $\cdot$ Fix $\omega\in E_{\mathbb{C}^N}$ and let $\mu_0$ be a minimizing measure for $E(\omega)$. If there exists $P$ such that $P\mu_0=\mu_0$ then $E_{h_Q}(\omega)\leq E(\omega)$.}

Define $E_{h_{Q,\Phi}}(\omega)=\inf_{\mu\in M_\omega(E)}\int h_{Q,\Phi}^r(\eta)\; d\mu(\eta)$, where $h_{Q,\Phi}^r(\rho)=h_Q(r\circ \Phi\rho)$. If $\Phi$ is a homogeneous quantum channel, e.g. with constant $p_i$, then we also have an upper bound. The proof follows from proposition \ref{homog_lem1}, item 4:

{\it $\cdot$ Fix $\omega\in E_{\mathbb{C}^N}$ and let $\mu_0$ be a minimizing measure for $E(\omega)$. Let $\Phi$ be a homogeneous quantum channel. If there exists $P$ such that $P\mu_0=\mu_0$ then $E_{h_Q}(\omega)\leq E(\omega)\leq E_{h_{Q,\Phi}}(\omega)$.}

\qee

\section{Projective metric}\label{sec_pmetric}

We recall some well-known facts about the projective metric. This will be needed in the subsequent sections. For general references, see \cite{birkhoff,wolf,viana}. First we provide general definitions, and later we specialize. Let $E$ be a vector space. A {\bf cone} is a subset $C\subset E-\{0\}$ such that $t>0$, $v\in C\Rightarrow tv\in C$. Define for $v_1,v_2\in C$,
\begin{equation}
\alpha(v_1,v_2):=\sup\{t>0: v_2-tv_1\in C\}, \;\;\;\; \beta(v_1,v_2):=\inf\{s> 0: sv_1-v_2\in C\}
\end{equation}
We have $\alpha(v_1,v_2)\leq\beta(v_1,v_2)$. Also $\alpha(v_1,v_2)<+\infty$ and $\beta(v_1,v_2)>0$. Define
\begin{equation}
\theta(v_1,v_2):=\log\frac{\beta(v_1,v_2)}{\alpha(v_1,v_2)}
\end{equation}
We call $\theta(\cdot,\cdot)$ the {\bf projective metric} associated to the convex cone $C$. Let $C_1\subset C_2$ be two convex cones in $E$. Let $\alpha_i,\beta_i,\theta_i$ the corresponding objects, $i=1,2$. Then $\alpha_1(v_1,v_2)\leq\alpha_2(v_1,v_2)$, $\beta_1(v_1,v_2)\geq\beta_2(v_1,v_2)$ and so $\theta_1(v_1,v_2)\geq\theta_2(v_1,v_2)$, for all $v_1,v_2\in C_1\subset C_2$.

\bigskip

More generally, let $E_1, E_2$ be vector spaces with $C_i\subset E_i$, $i=1,2$ be convex cones. Define $T:E_1\to E_2$ a linear operator such that $T(C_1)\subset C_2$. Then $\theta_1(v_1,v_2)\geq \theta_2(T(v_1),T(v_2))$. We have the following:
\begin{lemma}\label{exp_bas1}
If $D=\sup\{\theta_2(T(v_1),T(v_2)): v_1, v_2\in C_1\}<+\infty$ then
\begin{equation}
\theta_2(T(v_1),T(v_2))\leq (1-e^{-D})\theta_1(v_1,v_2)
\end{equation}
\end{lemma}

The value $D$ in the above lemma is called {\bf projective diameter} of T.
Such number is the main point of analysis when dealing with the projective metric. In the work \cite{wolf}, the authors show that if the projective diameter of the operator is finite, one obtains a nontrivial contraction of the trace norm (eq. (\ref{tanheq})). In our setting, the finiteness of the diameter will allow us to establish the exponential mixing property.

\bigskip

From now on we assume $E=C(D(\mathbb{C}^N);\mathbb{R})$, the space of continuous functions on $D(\mathbb{C}^N)$. Fix $\delta_0>0$, $a>0$ and $\nu>0$. Define the following convex cone of continuous functions:
\begin{equation}
C(a,\nu):=\{\psi\in E:\psi(\rho)>0, d(\rho_1,\rho_2)\leq \delta_0 \Rightarrow \psi(\rho_1)\leq exp(ad(\rho_1,\rho_2)^\nu)\psi(\rho_2)\}
\end{equation}
We impose no restriction on the metric $d$ so we can choose, for instance, the metric induced by the trace norm, $\Vert A\Vert_{tr}:=tr(\sqrt{A^*A})$ or Frobenius norm, $\Vert A\Vert:=\sqrt{\sum_{i,j}|A_{ij}|^2}$, for matrices. Define
\begin{equation}
C_+:=\{\psi\in E:\psi(\rho)>0, \forall \rho\in D(\mathbb{C}^N)\}
\end{equation}
The projective metric $\theta_+$ associated to $C_+$ is
\begin{equation}
\theta_+(\psi_1,\psi_2):=\log(\beta_+(\psi_1,\psi_2)/\alpha_+(\psi_1,\psi_2)),
\end{equation}
where
\begin{equation}
\alpha_+(\psi_1,\psi_2)=\inf\Big\{\frac{\psi_2}{\psi_1}(\rho):\rho\in D(\mathbb{C}^N)\Big\},\;\;\;\;\beta_+(\psi_1,\psi_2)=\sup\Big\{\frac{\psi_2}{\psi_1}(\rho):\rho\in D(\mathbb{C}^N)\Big\}
\end{equation}
Therefore,
\begin{equation}
\theta_+(\varphi_1,\varphi_2)=\log\sup\Big\{\frac{\varphi_2(\rho)\varphi_1(\eta)}{\varphi_1(\rho)\varphi_2(\eta)}: \rho,\eta\in D(\mathbb{C}^N)\Big\}.
\end{equation}

\section{Exponential decay of correlations}\label{mixingsection}

In this section we consider the transfer operator
\begin{equation}\label{topec}
T\varphi(\rho)=\sum_{i=1}^k\varphi(F_i(\rho)),
\end{equation}
with $F_i(\rho)=p_iU_i\rho U_i^*$, $U_i$ unitary, $p_i\in(0,1)$. We have the following:
\begin{lemma}\label{lema1a}
There is $\lambda_1<1$ such that $T(C(a,\nu))\subset C(\lambda_1a,\nu)$ for a sufficiently large $a>0$.
\end{lemma}
{\bf Proof.} We have
$$T\psi(\rho_1)=\sum_{i=1}^n \psi(F_i(\rho_1))\leq \sum_i exp(ad(F_i(\rho_1),F_i(\rho_2))^\nu)\psi(F_i(\rho_2))\leq$$
$$\leq exp(a\lambda_1d(\rho_1,\rho_2)^\nu)\sum_i \psi(F_i(y_2))=exp(a\lambda_1 d(\rho_1,\rho_2)^\nu)L\psi(\rho_2),$$
where $\lambda_1$ exists, due to the strict contractivity of the $F_i$.
\qed

\begin{lemma}\label{fdiam_lemma} (Finite diameter). $D_1=\sup\{\theta(\varphi_1,\varphi_2):\varphi_1,\varphi_2\in C(\lambda_1a,\nu)\}$ is finite, for all $a>0$, $\nu>0$, $\lambda_1<1$.
\end{lemma}
{\bf Proof.} See \cite{viana}.
\qed

Before we prove the existence of a fixed point, we state a lemma which might be of independent interest.

\begin{lemma}
Let $T$ be the transfer operator given by (\ref{topec}) on $C(D(\mathbb{C}^N))$. If there is $\lambda>0$ and $\varphi>0$, $\varphi\in C(D(\mathbb{C}^N))$ such that $T\varphi=\lambda \varphi$ then $\lambda=k$.
\end{lemma}
{\bf Proof.} The main idea comes from a result in \cite{jiang}. Let $m=\min_{\rho\in D(\mathbb{C}^N)}\varphi(\rho)$, $M=\max_{\rho\in D(\mathbb{C}^N)}\varphi(\rho)$. Then
$$0<\frac{m}{M}\leq \frac{\varphi(\rho)}{M}=\frac{\lambda^{-n}T^n\varphi}{M}=\lambda^{-n}T^n(\varphi/M)\leq \lambda^{-n}T^n 1=k^n\lambda^{-n}$$
In a similar way, we get $k^n\lambda^{-n}\leq M/m$. Hence
$$\frac{m}{M}\leq \frac{k^n}{\lambda^{n}}\leq \frac{M}{m}$$
Since the above holds for all $n$, we conclude that $\lambda=k$.

\qed

\begin{pro}
There exists a fixed point $\varphi_0$ for $T\varphi(\rho)=\sum_{i=1}^k\varphi(F_i(\rho))$.
\end{pro}
{\bf Proof.} First note that the spectral radius of $T$ is $r(T):=\lim_n\Vert T^n1\Vert^{1/n}=k$. Let $J=(j_1,j_2,\dots, j_m)$, $1\leq j_i\leq m$ and define
\begin{equation}
F_J:=F_{j_1}\circ F_{j_2}\circ\cdots\circ F_{j_m}
\end{equation}
\begin{equation}
\gamma_J(\rho):=\sup_{\rho\neq\eta}\frac{\Vert F_J(\rho)-F_J(\eta)\Vert}{\Vert\rho-\eta\Vert}
\end{equation}
Above, $\Vert\cdot\Vert$ is the Frobenius norm. By noting that $\Vert\cdot\Vert\leq\Vert\cdot\Vert_{tr}\leq \sqrt{N}\Vert\cdot\Vert$ we have
\begin{equation}
\frac{\Vert F_J(\rho)-F_J(\eta)\Vert}{\Vert\rho-\eta\Vert}\leq \frac{\sqrt{N}\Vert F_J(\rho)-F_J(\eta)\Vert_{tr}}{\Vert\rho-\eta\Vert_{tr}}\leq C_J\sqrt{N}
\end{equation}
for some $C_J<1$, which exists due to the strict contractivity of the $F_i$ mappings. Note that $C_J\to 0$ as $|J|=m\to\infty$. So we see that the inequality
\begin{equation}
\sup_{\rho\in D(\mathbb{C}^N)}\sum_{|J|=m} \gamma_J(\rho)<r(T)^m=k^m
\end{equation}
holds for $m$ sufficiently large. By \cite{jiang}, Theorem 4.5, we conclude that $T$ admits a fixed point, which we call $\varphi_0$.

\qed

Define $\mu_0=\varphi_0m$, where $m$ is some predefined measure. This will be used in the propositions that follow. In proofs of exponential decay for expansive dynamics, $m$ is Lebesgue measure and the invariance of $\mu_0$ under a (different) transfer operator is used for proving results related to ergodicity and stochastic stability, but we will use it in our setting so we can preserve the structure of the original proof.

\begin{lemma}
For $\varphi\in C(\lambda_1a,\nu)$, and the transfer operator given in (\ref{topec}) we have:
\begin{enumerate}
\item $\theta_+(T^n\varphi,\varphi_0)\leq \theta(T^n\varphi,\varphi_0)\leq\theta(\varphi,\varphi_0)\Lambda_1^n\leq D_1\Lambda_1^n$

\item $\sup\vert T^n\varphi-\varphi_0\vert\leq R_1(exp(D_1\Lambda_1^n)-1)\leq R_2\Lambda_1^n$,

\end{enumerate}
for some constants $R_1$ and $R_2$ and every $n\geq 1$.
\end{lemma}
{\bf Proof.} 1) The first inequality is due to the fact that on the LHS a supremum is being taken on a smaller set. The second inequality follows from lemmas \ref{exp_bas1}, \ref{fdiam_lemma} and the fact that $\varphi_0$ is fixed for $T$. The third inequality follows from the definition of $D_1$ (lemma \ref{fdiam_lemma}). For the proof of 2), note that the second inequality clearly holds for some constant $R_2>0$ and every $n\geq 1$. The first inequality is true because if $\{\varphi_l\}$ is a normalized Cauchy sequence and given any $\epsilon>0$, there exists $N\geq 1$ such that for all $k,l\geq N$,
\begin{equation}
\sup\frac{\varphi_k}{\varphi_l}\leq e^{\epsilon}
\end{equation}
Then
\begin{equation}
\sup |\varphi_k-\varphi_l|\leq\sup |\varphi_l|\sup\Big|\frac{\varphi_k}{\varphi_l}-1\Big|\leq C(e^\epsilon-1).
\end{equation}

\qed

In order to prove the exponential decay, we still need the following fact. The proof is very similar to the one for the equality $\langle T_b\varphi,\mu\rangle=\langle\varphi,P_b\mu\rangle$, described for the barycentric transfer operator in section \ref{snotpre}.

\begin{lemma}\label{escapelemma}
Given the transfer operator $T$ in (\ref{topec}), we have for any integrable functions $\varphi$, $\psi$,
\begin{equation}
\int (T\varphi)\psi \;d\mu=\int \varphi\;  dP_e\mu_\psi,
\end{equation}
for any fixed $\mu$, where $d\mu_\psi:=\psi\; d\mu$ and $P_e\mu(B)=\sum_i\int_{F_i^{-1}(B)} d\mu$.
\end{lemma}

\begin{theorem}
Given the transfer operator in (\ref{topec}), a H\"older function $\varphi$ and $\psi\in L^1(m)$ a function on $D(\mathbb{C}^N)$, there is $K=K(\varphi,\psi)>0$ such that
\begin{equation}\label{embeq}
\Bigg\arrowvert\int \psi(T^n\varphi)\; dm - \int\psi\; d\mu_0\int\varphi \;dm\Bigg\arrowvert\leq K\Lambda_1^n, \;\; n\geq 0,
\end{equation}
where $\Lambda_1=1-e^{-D_1}<1$, and $D_1=\sup\{\theta(\varphi_1,\varphi_2):\varphi_1,\varphi_2\in C(\lambda_1a,\nu)\}$, $a>0$, $\nu >0$.
\end{theorem}
{\bf Proof.} First assume that $\varphi\in C(\lambda_1a,\nu)$. Without loss of generality, assume $\int \varphi \;dm=1$. Denote $\Vert\psi\Vert_1=\int \vert\psi\vert\; d\mu_0$, then
$$\Bigg\vert\int(T^n\varphi)\psi\;dm-\int\psi\;d\mu_0\Bigg\vert=\Bigg\vert\int \psi\Big(\frac{T^n\varphi}{\varphi_0}-1\Big)d\mu_0\Bigg\vert\leq \Bigg\Vert \frac{T^n\varphi}{\varphi_0}-1\Bigg\Vert_0\Vert\psi\Vert_1$$
\begin{equation}
\leq R_1'(exp(D_1\Lambda_1^n)-1)\Vert\psi\Vert_1\leq R_2'\Vert\psi\Vert_1\Lambda_1^n,
\end{equation}
recall that $\varphi_0>0$. For a general $\nu$-H\"older $\varphi$, let $A>0$ such that $\varphi$ is $(A,\nu)$-H\"older, and for $B>0$ write
\begin{equation}
\varphi=\varphi_B^+-\varphi_B^{-},\;\;\; \varphi_B^{\pm}=\frac{1}{2}(|\varphi|\pm\varphi)+B
\end{equation}
If we take $B=(A/\lambda_1a)$ we get $\varphi_B^{\pm}\in C(\lambda_1a,\,\nu)$ so the proposition holds for $\varphi_B^{\pm}$, so it holds for $\varphi$ by linearity (on equation (\ref{embeq}) and using lemma \ref{escapelemma}).

\qed

\begin{cor}
Given $\nu$-H\"older continuous functions $\varphi, \psi$, there exists $K=K(\varphi,\psi)>0$ such that
\begin{equation}\label{embeq2}
\Bigg\arrowvert\int \psi(T^n\varphi)\; dm - \int\psi\varphi_0\; dm\int\varphi \;dm\Bigg\arrowvert\leq K\Lambda_1^n, \;\; n\geq 0.
\end{equation}

\end{cor}

\section{Conclusion}

Part of our work was motivated by the classification theorem described in \cite{petulante} and the asymptotic analysis made in \cite{novotny}. We ask the question: is there a transfer operator associated to a CPT map $\Phi$ which gives us some information on the eigenspaces of $\Phi$? It is tempting to conjecture that if a transfer operator has a unique fixed point, then by Ruelle's theorem and the barycenter theorem $\Phi$ has a unique fixed point (i.e., $\dim \ker(\Phi-I)=1$). This conjecture is false for the transfer operator considered in section \ref{mixingsection}: take for instance the phase-flip channel on single qubits which possesses 1 as the only eigenvalue on the unit circle and $\dim \ker(\Phi-I)=2$, see \cite{petulante}. Yet such channel is mixed-unitary, so Ruelle's theorem is valid, by
theorem 4.5 in \cite{jiang}. Clearly, in order to determine the number of fixed points, one could just calculate eigenvalue and eigenvectors of the associated matrix, but a description via transfer operators could provide new insights on CPT maps.

\bigskip

\section{Acknowledgments}

The author would like to thank A. Baraviera for helpful conversations on topics related to this work.

\end{document}